\pdfoutput=1
\documentclass[aps,prc,twocolumn,floatfix,superscriptaddress,nofootinbib,preprintnumbers,longbibliography,amsmath,amsthm,amssymb]{revtex4-2}
\usepackage{graphicx}

\usepackage{amsfonts}
\usepackage{color}
\usepackage[colorlinks=true,linkcolor=blue,citecolor=blue]{hyperref}
\usepackage{dcolumn}
\usepackage{bm}
\usepackage{braket}
\usepackage[T1]{fontenc}

\newcommand{\be}{\begin{equation}}
\newcommand{\ee}{\end{equation}}
\newcommand{\dd}{\mathrm{d}}
\newcommand{\br}{\boldsymbol{r}}

\newcommand{\ii}{\mathrm{i}}
\newcommand{\bvec}[1]{\mbox{\boldmath $#1$}}

\begin{document}
\allowdisplaybreaks[1]
\title{Super- and hyper-deformation in $^{60}$Zn, $^{62}$Zn, and $^{64}$Ge at high spins
}

\author{Kenichi Yoshida}
\email[E-mail: ]{kyoshida@ruby.scphys.kyoto-u.ac.jp}
\affiliation{Department of Physics, Kyoto University, Kyoto, 606-8502, Japan}

\preprint{KUNS-2890}
\date{\today}

\begin{abstract}
\begin{description}
\item[Background] 
The observation of the superdeformed (SD) bands in $^{60,62}$Zn indicates that the particle number 30 is a magic particle number, 
where two and four neutron single-particles are considered to be promoted to the intruder $1g_{9/2}$ shell. 
However, the SD-yrast band in $^{62}$Zn is assigned negative parity. 
\item[Purpose] I investigate various SD configurations in the rapidly rotating $^{60,62}$Zn and $^{64}$Ge, 
and attempt elucidating the different roles of the energy gaps at particle numbers 30 and 32. 
\item[Method] I employ a nuclear energy-density functional (EDF) method: the configuration-constrained cranked Skyrme--Kohn--Sham approach is 
used to describe the rotational bands near the yrast line.
\item[Results]  
The negative-parity SD bands appear higher in energy than the positive-parity SD-yrast band in $^{60}$Zn by about 4 MeV, 
which is indicative of the SD doubly-magic nucleus. 
However, the energy gap in $^{64}$Ge is smaller $\sim$ 2--3 MeV, though 
the quadrupole deformation of the SD states in $^{64}$Ge is greater than that of $^{60}$Zn. 
The present calculation predicts the occurrence of the hyperdeformed state in $^{60}$Zn and $^{64}$Ge at a high rotational frequency 
$\sim 2.0$ MeV$/\hbar$ due to the occupation of the $h_{11/2}$ shell.
\item[Conclusions]
An SD-shell gap at particle number 30 and 32 appears at different deformations and the energy gap at particle number 32 is low, 
which make the SD structures of $^{62}$Zn unique, where the negative-parity SD states appear lower in energy than the positive-parity one.
\end{description}
\end{abstract}

\maketitle

\section{Introduction}\label{intro}

A recent expansion of accelerator facilities, development of $\gamma$-ray and particle detector systems, 
and progress in techniques of 
nuclear spectroscopy and theoretical many-body calculation 
have significantly advanced the study of exotic nuclei. 
An extreme is nuclei in high-spin and highly-elongated states. 
Since the discovery of the superdeformed (SD) band in $^{152}$Dy~\cite{twi86}, 
the SD bands have been observed up to high spins in various mass regions~\cite{sin02},  
including the lighter mass region $A\sim60$~\cite{sve97,rev02,rud99,jon08,rud10,rud06,jon09,rud98,and00,and02,and03,and08,sve99,yu99,and09,yu00}.

The yrast spectroscopy of light nuclei gives a unique opportunity to investigate the microscopic mechanism 
for the occurrence of the SD band and the possibility of the hyperdeformed (HD) band 
because the single-particle density of state around the Fermi levels is low so that 
one can study in detail the deformed shell structures responsible for the SD and HD bands~\cite{ina02,ray16,sak20}.

$N \simeq Z$ nuclei in the $A \sim 60$ mass region go through the evolution of shapes with increasing nuclear spin. 
The low-spin structures are dominated by the single-particle and collective excitations generated by the 
valence particles outside the doubly-magic $^{56}$Ni core: the $\mathcal{N}=3$ $pf$ shell.
At intermediate and high spins, the particle--hole excitations across the magic number 28 and the occupation of 
the $\mathcal{N}=4$ intruder $1g_{9/2}$ shell characterize the change from well-deformed (WD) to SD shapes~\cite{sve97}.

Thanks to an SD-shell gap at the particle numbers $N,Z=30,32$~\cite{dud87}, 
the SD bands have been observed in $^{60,62}$Zn~\cite{sve99,sve97}, 
in which two and four neutrons are expected to occupy the deformation-driving $g_{9/2}$ shell. 
The SD band in $^{60}$Zn is well described using the 
configuration-dependent cranked Nilsson--Strutinsky (CNS) method and 
the cranked relativistic mean-field (CRMF) method for the configuration involving two neutrons in the $g_{9/2}$ shell~\cite{sve99}.
However, the theoretical calculations have ruled out 
the configuration involving four neutrons in the $g_{9/2}$ shell 
for the observed SD band, called SD1, in $^{62}$Zn~\cite{mad98}. 

Combined data for $^{62}$Zn 
from four experiments have disclosed two new SD bands, SD2 and SD3, in Ref.~\cite{gel09} and two more SD bands, SD4 and SD5, in Ref.~\cite{gel12}. 
It then turned out that the naturally-expected SD band appears at relatively high energy as SD3, 
as pointed out theoretically. 
The rich information on high-spin structures in $^{62}$Zn provides insights into how the SD states emerge and evolve 
in varying particle numbers and spins.

The present work aims to investigate the various SD configurations in $^{60,62}$Zn and $^{64}$Ge, 
compare with the available experimental data, 
and elucidate the microscopic mechanism for the change of the SD structures from $^{60}$Zn to $^{62}$Zn. 
I use a nuclear energy-density-functional (EDF) method: 
a theoretical model being capable of handling nuclides with arbitrary mass numbers~\cite{ben03,nak16}. 
Then I find that 
the SD states in $^{64}$Ge have a larger deformation than $^{60}$Zn whereas 
the SD-shell gap in $^{64}$Ge is lower than that in $^{60}$Zn.
Furthermore, the possible appearance of the HD band in $^{60,62}$Zn and $^{64}$Ge is discussed. 

This paper is organized in the following way: 
the theoretical framework for describing the SD/HD bands is given in Sec.~\ref{model} and 
some details of the numerical procedures are also given; 
Sec.~\ref{result} is devoted to the numerical results and discussion based on the model calculation; 
the SD bands in $^{60}$Zn, $^{62}$Zn, and $^{64}$Ge are investigated in Sec.~\ref{60Zn}, Sec.~\ref{62Zn}, and Sec.~\ref{64Ge}, 
respectively; the HD bands are discussed in Sec.~\ref{hyper}.  
Then, a summary is given in Sec.~\ref{summary}.

\section{Theoretical model}\label{model}

\subsection{Cranked Skyrme--Khon--Sham calculations}
Since the details of the formalism can be found in Ref.~\cite{sak20}, 
here I briefly recapitulate the basic equations relevant to the present study. 
In the framework of the nuclear EDF method I employ, I solve 
the cranked Skyrme--Khon--Sham (SKS) equation~\cite{bon87} obtained by 
\begin{equation}
\delta (E[\rho] - \omega_\mathrm{rot}\braket{ \hat{J}_z})=0,
\label{eq:cranking}
\end{equation}
where $E[\rho]$ is a nuclear EDF, 
$\omega_\mathrm{rot}$ and $\hat{J}_z$ mean the rotational frequency and 
the $z$-component of angular momentum operator, 
and the bracket denotes the expectation value with respect to the Slater determinant 
given by the occupied single-particle KS orbitals for a given $\omega_\mathrm{rot}$. 
I define the $z$-axis as a quantization axis of the intrinsic spin and 
consider the system rotating uniformly about the $z$-axis. 
I take the natural units: $\hbar=c=1$. 

In the present calculation, a Skyrme-type EDF is employed as the nuclear EDF~\cite{dob95}:  
\begin{align}
E&=E_{\rm kin} + E_{\rm Sky} + E_{\rm Coul}, \\ 
E_{\rm Sky}&=\int \dd \br \mathcal{H}_{\rm Sky}(\br), \label{eq:Skyrme_EDF}\\
\mathcal{H}_{\rm Sky}(\br)&=\sum_{t=0,1}[ C_t^{\rho}\rho^2_t + C_t^s \boldsymbol{s}^2_t + C^{\Delta \rho}_t \rho_t \Delta \rho_t \notag \\
&+ C^\tau_t(\rho_t \tau_t - \boldsymbol{j}^2_t) + C^{\nabla J}_t(\rho_t \nabla \cdot \boldsymbol{J}_t 
+ \boldsymbol{s}_t \cdot \nabla \times \boldsymbol{j}_t) ].
\end{align}

\subsection{Numerical procedures}

In solving the cranked-KS equation, 
I impose the reflection symmetry about the $(x, y)$-, $(y,z)$- and $(z, x)$-planes. 
One can therefore construct the simultaneous engenfunctions of the parity transformation $\hat{P}$ 
and the $\pi$ rotation about the $z$-axis $\hat{R}_z=e^{-\ii \pi \hat{j}_z}$:
\begin{align}
    \hat{P}\phi_k&=\mathfrak{p}_k\phi_k, \\
    \hat{R}_z\phi_k&=r_k\phi_k,
\end{align}
besides the cranked-KS equation
\begin{equation}
\hat{h}^\prime \phi_k = \epsilon_k \phi_k, 
\label{eq:cSKS}
\end{equation}
with the single-particle Hamiltonian, or the Routhian for $\omega_\mathrm{rot} \ne 0$, namely 
$\hat{h}^\prime=\frac{\delta E}{\delta \rho} - \omega_\mathrm{rot}\hat{j}_z$. 
The eigenvalues $\mathfrak{p}_k$ $(=\pm 1)$ and $r_k$ ($=\pm \ii$) are called 
the parity and $z$-signature, respectively. Hereafter, I simply call the latter signature. 
One can introduce the signature exponent quantum number $\alpha$ ($=\pm 1/2$) by $r \equiv e^{-\ii \pi \alpha}$. 
The signature exponent $\alpha$ is useful when comparing with the experimental data through 
the relation $\alpha=I \mod 2$, where $I$ is the total nuclear spin~\cite{voi83}.

I solve Eq.~(\ref{eq:cSKS}) by diagonalizing the single-particle Routhian $\hat{h}^\prime$
in the three-dimensional Cartesian-mesh representation with the box boundary condition. 
Thanks to the reflection symmetries, I have only to consider the octant region explicitly in space 
with $x\ge0$, $y\ge0$, and $z\ge0$; see Ref.~\cite{oga09} for details.
I use a 3D lattice mesh $x_i=ih-h/2, y_j=jh-h/2, z_k=kh-h/2 \ \  (i,j,k=1,2,\cdots)$ with a mesh size $h=1.0$ fm and 12 points for each direction. 
The differential operators are represented by the use of the 9-point formula of the finite difference method.
For diagonalization of the Routhian, I use the LAPACK {\sc dsyevx} subroutine~\cite{LAPACK}. 
A modified Broyden's method~\cite{bar08} is utilized to calculate new densities during the selfconsistent iteration. 

To describe various types of rotational bands under the energy variation, 
the Slater determinantal states are constructed by imposing the configuration of the single-particle KS orbitals. 
Since the parity and signature are a good quantum number, 
and the pairing correlations are not included in the present calculation, 
the intrinsic configurations of interest can be
described by the occupation number of particle $n$ for the orbitals specified by the quantum number $(\mathfrak{p}, \alpha)$; 
$[n_{(+1, +1/2)} n_{(+1,-1/2)}n_{(-1, +1/2)}n_{(-1, -1/2)}]_q$ for $q=\pi$ (proton) and $\nu$ (neutron) 
similarly to the cranking calculation code {\sc hfodd}~\cite{HFODD}. 

The structure of the configurations is investigated by looking at the deformation. 
Thus, the multipole moments are introduced as  
\begin{equation}
    \alpha_{lm}=\frac{4\pi}{3A\bar{R}^l}\int d^3r\ r^lX_{lm}(\hat{r})\rho(\bvec{r}),
  \label{eq:MultMome} 
\end{equation}
where $\rho(\bvec{r})$ is the particle density, 
$\bar{R}=\sqrt{\tfrac{5}{3A}\int d^3r\ r^2\rho(\bvec{r})}$, and 
$X_{lm}$ are real basis of the spherical harmonics: 
\begin{equation}
  X_{lm}(\hat{r})=
  \begin{cases}
    Y_{l0}(\hat{r}) & (m=0)\\
    \tfrac{1}{\sqrt{2}}[Y_{l,-m}(\hat{r})+Y^\ast_{l,-m}(\hat{r})] & (m>0) \\
    \tfrac{1}{\sqrt{2}i}(-1)^m[Y_{lm}(\hat{r})-Y^\ast_{lm}(\hat{r})] & (m<0).
  \end{cases}
\end{equation}
I then define the quadrupole deformation parameter $\beta$ and the triaxial deformation parameter $\gamma$~\cite{nil95} by
\begin{equation}
  \alpha_{20}=\beta\cos\gamma,\ \ \ \alpha_{22}=-\beta\sin\gamma.
\end{equation}

\section{results and discussion}\label{result}

The $j$ shells characterizing the highly-deformed structures in nuclei with $A \sim 60$ 
are the $\mathcal{N}=3$ high-$j$ $1f_{7/2}$, 
low-$j$ $pf$-shells, and the $\mathcal{N}=4$ intruder $1g_{9/2}$ shell. 
Thus, the configurations can be specified as
\begin{align}
\pi[&(1f_{7/2})^{-p_1}(pf)^{p_3}(1g_{9/2})^{p_2}] \notag \\
&\otimes \nu[(1f_{7/2})^{-n_1}(pf)^{n_3}(1g_{9/2})^{n_2}]
\end{align}
relative to the $^{56}$Ni core, or as $[p_1 p_2, n_1 n_2]$ simply. 
It is noticed that the Nilsson-type deformed wave functions at large deformations 
are not simply represented by the orbitals within the single major shell, but 
are expressed by the linear combination of 
the spherical orbitals with admixture of different major shells of $\Delta \mathcal{N}=2$.
Therefore, the classification of the highly-deformed bands according to the spherical shells is just for a practical convenience. 
Furthermore, one cannot distinguish the hole states in a $2p$ and $1f$ orbital because they have the same parity, 
as summarized in Table~\ref{tab:config}.

\begin{table*}[t]
\caption{\label{tab:config} 
Investigated configurations of SD and quasi-HD (involving one particle in the $1h_{11/2}$ orbital) 
in terms of the spherical shells and in the present calculation scheme. 
The ``($+$)'' or ``($-$)'' labels specify the signature of an odd number of protons or neutrons 
in the $pf$ or $g_{9/2}$ orbitals with ($+$) for $\alpha=+1/2$ and ($-$) for $\alpha=-1/2$. }
\begin{ruledtabular}
\begin{tabular}{ccccccc}
\multicolumn{2}{c}{$^{60}$Zn} &  \multicolumn{2}{c}{$^{62}$Zn} &  \multicolumn{2}{c}{$^{64}$Ge} \\
\cline{1-2} \cline{3-4} \cline{5-6}
$[7788]_\pi $ & $[00,$       & $[7788]_\pi$  & $[00,$                                        & $[9977]_\pi$ & $[24,$ or $[24(1),$   \\
$[8877]_\pi$ & $[22,$        & $[8877]_\pi$  & $[22,$ or $[22(1),$                           & $[8978]_\pi$ & $[2(+)3(-)$, or  $[2(+)3(-)(1),$ \\
$[9867]_\pi$ & $[1(-)1(+)$,  &  $[7799]_\nu$  & $00]$                                        & $[8987]_\pi$ & $[2(-)3(-),$ \\                   
$[8967]_\pi$ & $[1(-)1(-)$,  & $[8888]_\nu$  & $02]$ or $12]$ or $22]$ or $22(1)]$           & $[9878]_\pi$ & $[2(+)3(+),$ or  $[2(+)3(+)(1),$\\
$[9876]_\pi$ & $[1(+)1(+)$,  & $[8978]_\nu$  & $1(+)3(-)]$ or $2(+)3(-)]$ or $[2(+)3(-)(1)]$ & $[9887]_\pi$ & $[2(-)3(+),$\\                    
$[8976]_\pi$ & $[1(+)1(-)$,  & $[8987]_\nu$  & $1(-)3(-)]$ or $2(-)3(-)]$                    & $[9977]_\nu$ & $24]$ or $24(1)]$  \\
 $[7788]_\nu$ & $00]$ & $[9878]_\nu$  & $1(+)3(+)]$ or $2(+)3(+)]$ or $[2(+)3(+)(1)]$        & $[8978]_\nu$ & $2(+)3(-)]$ or  $2(+)3(-)(1)]$ \\
 $[8877]_\nu$ & $22]$  & $[9887]_\nu$  & $1(-)3(+)]$ or $2(-)3(+)]$                          & $[8987]_\nu$ & $2(-)3(-)]$\\
$[9876]_\nu$ & $1(+)1(+)]$   & $[9977]_\nu$  & $24]$ or $24(1)]$                             & $[9878]_\nu$ & $2(+)3(+)]$ or  $2(+)3(+)(1)]$\\
$[8976]_\nu$ & $1(+)1(-)]$   &               &                                               & $[9887]_\nu$ & $2(-)3(+)]$\\                     
$[9867]_\nu$ & $1(-)1(+)]$   &               &                                               &              &                                \\
$[8967]_\nu$ & $1(-)1(-)]$   &               &&                                              
\end{tabular}
\end{ruledtabular}
\end{table*}

\subsection{Superdeformed states in $^{60}$Zn}\label{60Zn}

\begin{figure}[t]
\includegraphics[scale=0.49]{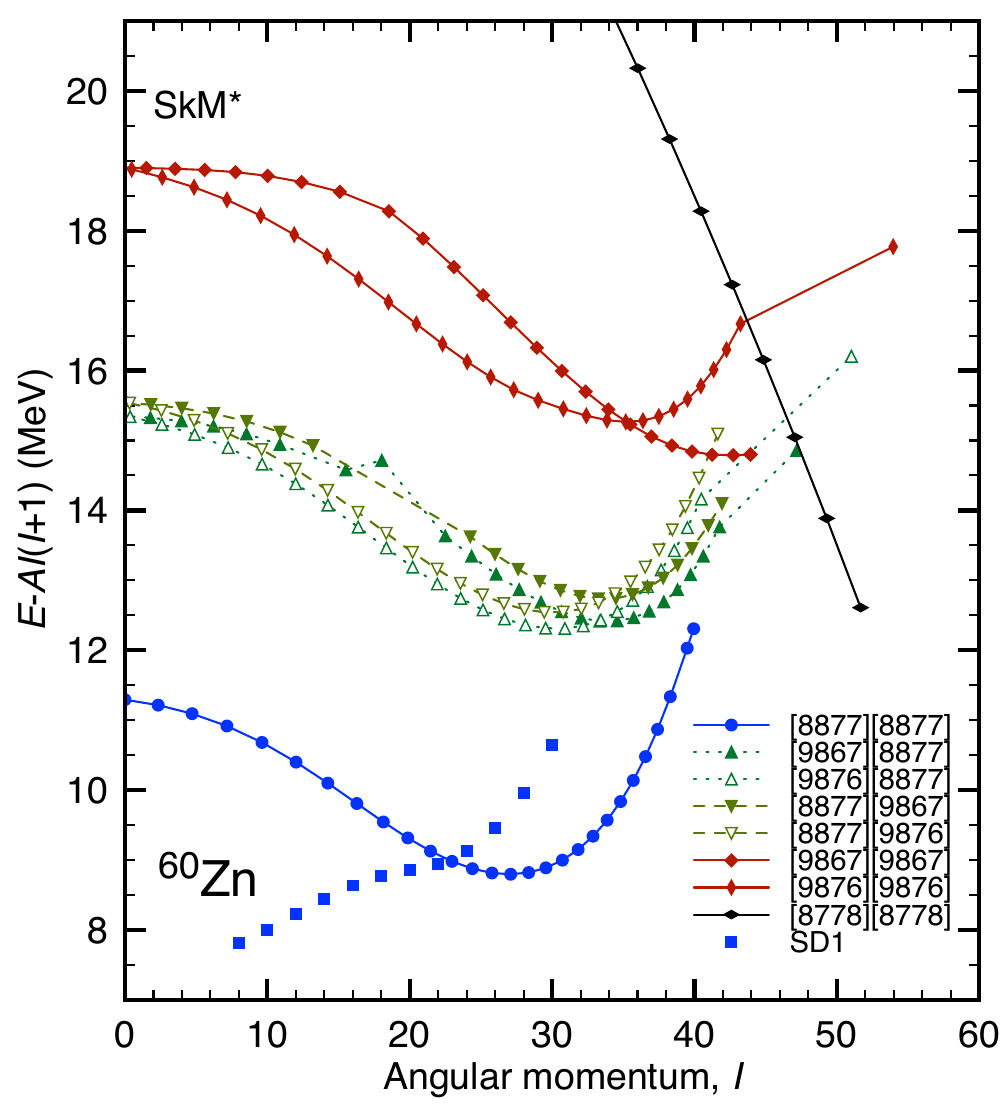}
\caption{\label{fig:60Zn_E}
Excitation energies of the rotational bands in $^{60}$Zn. 
Solid (dotted) lines represent positive (negative) parity, 
and closed (open) symbols denote signature $\alpha=0$ $(\alpha=+1)$. 
A smooth part $A I(I+1)$ is subtracted with an inertia parameter $A=0.025$ MeV in plotting the excitation energies. 
The subscript $\pi$ or $\nu$ to specify the configuration is omitted.
}
\end{figure}

\begin{figure}[t]
\includegraphics[scale=0.49]{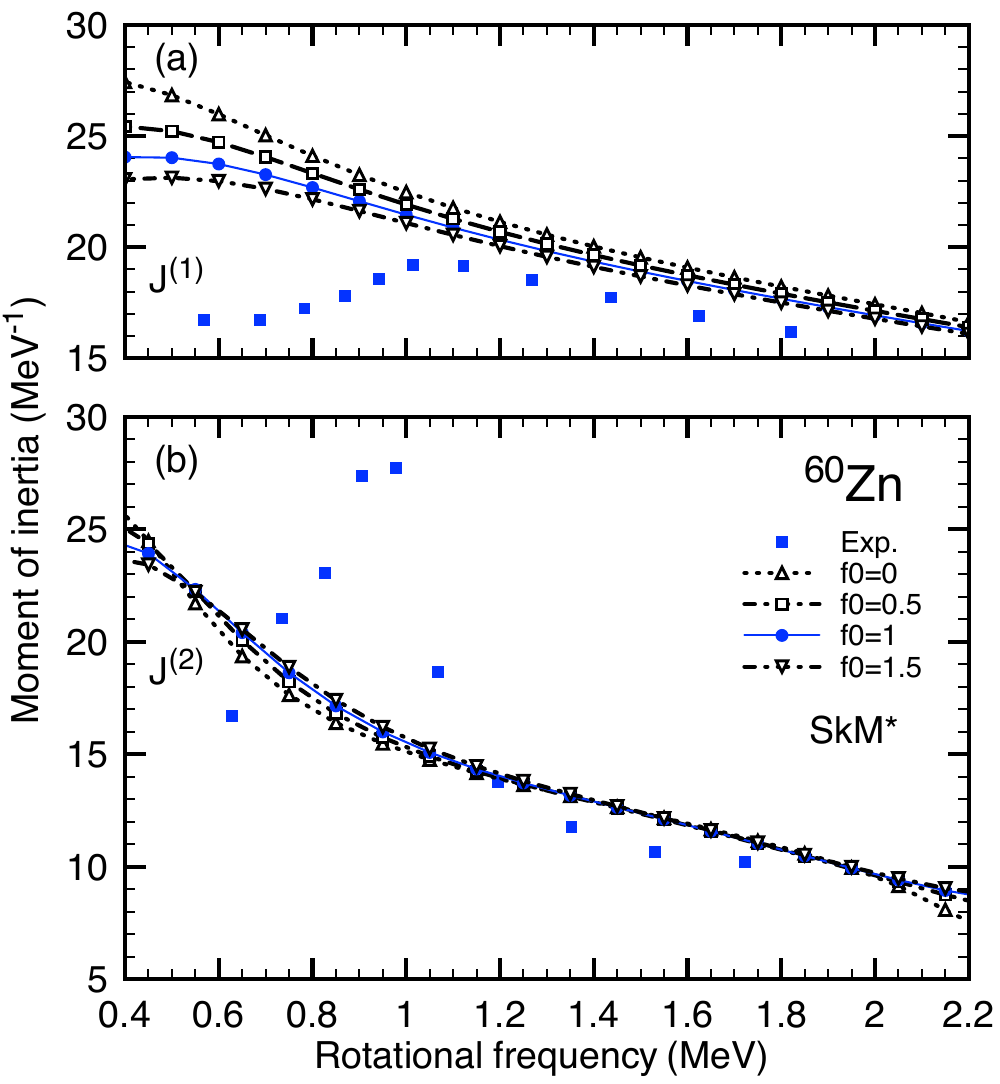}
\caption{\label{fig:60Zn_J}
(a) Kinematic and (b) dynamic moments of inertia of the SD band in $^{60}$Zn. 
Numerical results obtained by varying $f_0$ are also shown, 
where $C_0^s$ is multiplied by $f_0$.
}
\end{figure}

The normal-deformed state in $^{60}$Zn is represented as $[7788]_\pi [7788]_\nu$, 
in which two protons and two neutrons occupy the $[321]1/2$ orbital stemming from the $2p_{3/2}$ orbital; see, e.g., Fig.~1 of Ref.~\cite{yos05}.
With an increase in the prolate deformation, the $[303]7/2$ orbital stemming from the $1f_{7/2}$ shell grows in energy, 
and intersects with the $[310]1/2$ orbital originating from the $2p_{1/2}$ orbital. 
When the system is further deformed, an intruder orbital $[440]1/2$ coming down from the $1g_{9/2}$ shell crosses with the $[310]1/2$ orbital 
and an energy gap appears after the crossing, leading to an SD configuration represented as $[8877]_\pi [8877]_\nu$. 

The SD band `SD1' is assigned positive parity and comprises the states with $I=8\textrm{--}30$, 
thus $\alpha=0$~\cite{sve99}. 
Figure~\ref{fig:60Zn_E} shows the excitation energies relative to the reference energy $0.025 I(I+1)$. 
Here, the angular momentum $I$ is calculated as $\langle \hat{J}_z \rangle$. 
The ground state was obtained by solving the SKS--Bogoliubov equation~\cite{yos21c} with the use of the pairing 
energy in Ref.~\cite{yam09}. 
The configuration of SD1 is assigned $[22,22]$~\cite{sve99}, corresponding to the configuration 
involving two protons and two neutrons in the intruder $g_{9/2}$ shell and two-proton holes and two-neutron holes in the $f_{7/2}$ shell, 
as mentioned above. 
The present calculation overestimates the kinematic moments of inertia $\mathcal{J}^{(1)}$ in $\omega_{\rm rot}\lesssim 1$ MeV ($I \sim 20$). 
Experimentally, the enhancement in the dynamic moment of inertia $\mathcal{J}^{(2)}$ is observed 
around $\omega_{\rm rot}=0.9$ MeV ($I \sim 18$), as shown in Fig.~\ref{fig:60Zn_J}(b), 
which suggests the alignment of the $g_{9/2}$ protons and neutrons~\cite{sve99}.
Above $\omega_{\rm rot} \sim 1.1$ MeV ($I \sim 21$), the calculation reproduces the measured $\mathcal{J}^{(2)}$. 
The present calculation describes well 
the $\omega_{\rm rot}$ dependence of $\mathcal{J}^{(1)}$ as well above $\omega_{\rm rot} \sim 1.1$ MeV.

The deformation parameter of the SD band is obtained as $\beta=0.53$ at $\omega_{\rm rot}=0$ MeV, 
and decreases monotonically as increasing the rotational frequency: $\beta = 0.43$ at $\omega_{\rm rot}=1.4$ MeV ($I \sim 27$). 
The observed transition quadrupole moment in the spin range $I=12\text{--}27$ is $2.75 \pm 0.45$ $e{\rm b}$, corresponding 
to $\beta=0.47 \pm 0.07$~\cite{sve99}. 
The present calculation reproduces well the experimental value. 

The ``one-particle-one-hole (1p1h)'' and ``two-particle-two-hole (2p2h)'' excitations 
from the SD band may appear near the yrast line unless the SD shell-gap is developed. 
The SD shell-gap is generated by the gap between the $[440]1/2$ and $[310]1/2$ orbitals. 
Thus, there are four proton ``1p1h'' excitations from $[440]1/2(\alpha=\pm 1/2)$ to 
$[310]1/2(\alpha=\pm 1/2)$, labeled as $[8967]_\pi, [8976]_\pi, [9867]_\pi$, and $[9876]_\pi$,
while keeping the neutron configuration $[8877]_\nu$. 
Furthermore, there are four neutron ``1p1h'' excitations keeping the proton configuration $[8877]_\pi$. 
They are negative parity and correspond to the configuration involving one proton (neutron) and two neutrons (protons) in the $g_{9/2}$ shell. 
I show in Fig.~\ref{fig:60Zn_E} the energies of the favored configurations. 
A large gap $\sim 4$ MeV in the wide spin range can be seen. 
The ``2p2h'' configurations appear much higher in energy.
The positive-parity SD band is thus well apart from the excited bands.
This is supported by a strong population of the SD band experimentally~\cite{sve99}.
It is noticed that all the configurations under investigation reach the oblate deformation $\gamma=60^\circ$. 

At the end of this section, I investigate the uncertainty associated with the time-odd mean fields. 
In the Skyrme EDF (\ref{eq:Skyrme_EDF}), the coupling constants $C^s_t$ are determined 
solely by the time-odd observables. These are related to the 
Landau--Migdal parameters $g_0$ and $g_0^\prime$. 
Thus, the magnetic responses are sensitive to the values of $C_t^s$. 
The parameter $C_1^s$ has been adjusted to reproduce the experimentally observed 
Gamow--Teller (GT) resonance~\cite{gia81,ben02,roc12}. 
However, the experimental data for the isoscalar spin-flip excitations are too scarce~\cite{har01} 
to constrain the value of $C_0^s$. 
To see the uncertainty with respect to the coupling constants  $C_t^s$, 
I have performed the calculations by multiplying factors as $C_t^s \to f_t \times C_t^s$.

Figure~\ref{fig:60Zn_J} shows the kinematic and dynamic moments of inertia of the SD band. 
I found that the change in $f_1$ is negligibly small. 
This is rather reasonable because the density distributions of neutrons and protons are not 
very different in such an $N=Z$ nucleus, leading to a reduced contribution of the isovector density. 
Furthermore, the SkM* functional describes well the magnetic responses, 
such as the M1 and GT excitations in Ca and Ni isotopes~\cite{yos21} and 
the spin-dipole excitation in neutron-rich Ni isotopes~\cite{yos19}. 
Thus, the $C_1^s$ value has already been examined, and its change makes the description of the isovector spin-flip excitations worse. 
Here, I show only the results by varying $f_0$. 
In all the cases under investigation, the uncertainty is small. 
One sees that $\mathcal{J}^{(2)}$ is less sensitive to the values of $C_0^s$ whereas 
one can see a systematic trend in the calculated $\mathcal{J}^{(1)}$: 
the smaller the coupling constant, the higher the calculated $\mathcal{J}^{(1)}$. 
For the SkM* functional, the value of $C_0^s$ is 31.7 MeV fm$^3$ at the normal nuclear density. 
Then, the spin--spin interaction is repulsive for $S=1$ in the isoscalar channel. 
The spin alignment is thus favored for a weaker spin--spin interaction. 

\subsection{Superdeformed states in $^{62}$Zn}\label{62Zn}

The SD band `SD1' is assigned negative parity and comprises the states with $I=18\textrm{--}34$, 
thus $\alpha=0$~\cite{gel09}. 
`SD2' is considered as a signature partner of SD1 and covers a spin range of $17\textrm{--}35$~\cite{gel09}.
`SD3' is a positive-parity band covering $I=16\textrm{--}30$, thus $\alpha=0$~\cite{gel09}. 
In addition, `SD4' and `SD5' have been newly observed in Ref.~\cite{gel12}. 
These observed bands were analyzed using the configuration-dependent cranked Nilsson--Strutinsky (CNS) method~\cite{ben85,afa99}.
According to the CNS analysis, the configuration of 
SD1, SD2, SD3, and SD5 is considered as
$[22,2(-)3]$, $[22,2(+)3]$, $[22,24]$, and $[22,13]$, respectively~\cite{gel12}, whereas 
that of SD1, SD2, and SD5 is suggested as 
$[22,1(-)3]$, $[22,1(+)3]$, and $[22,2(+)3]$, respectively in a revised consideration~\cite{gel14}. 
The CRMF approach was applied earlier to investigate the SD1 band and assigned the configuration as $[22,24]$~\cite{afa98}.
However, the authors in Ref.~\cite{mad98} pointed out that the configuration [22,24] is not yrast and 
SD1 can be a negative-parity band 
based on the CRMF and the cranked Skyrme-KS methods.

In the present formalism, the ground state is specified as $[7788]_\pi [7799]_\nu$ with the quantum number 
$(\mathfrak{p}, \alpha)=(+1,0)$. 
The well-deformed (WD) states involve two protons and two neutrons in the $1g_{9/2}$ shell, corresponding to $[8877]_\pi [8888]_\nu$ with 
$(+1,0)$. 
The SD bands involve one or two more neutrons in the $1g_{9/2}$ shell, corresponding to 
$[8877]_\pi [8978]_\nu$ $(+1,+1)$,  
$[8877]_\pi [8987]_\nu$ $(-1,0)$, 
$[8877]_\pi [9878]_\nu$ $(-1,0)$,  
$[8877]_\pi [9887]_\nu$ $(-1,+1)$, 
or $[8877]_\pi [9977]_\nu$ $(+1,0)$. 
\footnote{
In Ref.~\cite{mad98}, these configurations were labeled as SD-D3, SD-D4, SD-D1, SD-D2, and SD-C, respectively. 
}
To reiterate, in the present calculation, the configurations are specified not by the spherical orbitals but by the parity. 
Therefore, the mixing among $\Delta \mathcal{N}=2$ is automatically included in the energy minimization. 
Instead, one cannot describe, for example, the $[22,1(-)3(+)]$ and $[22,2(-)3(+)]$ configurations separately 
because they belong to the specified same configuration of neutrons as $[9887]_{\nu}$
and one has the level crossing of the $2p$ and $1f$ orbitals in varying the rotational frequency.

\begin{figure}[t]
\includegraphics[scale=0.49]{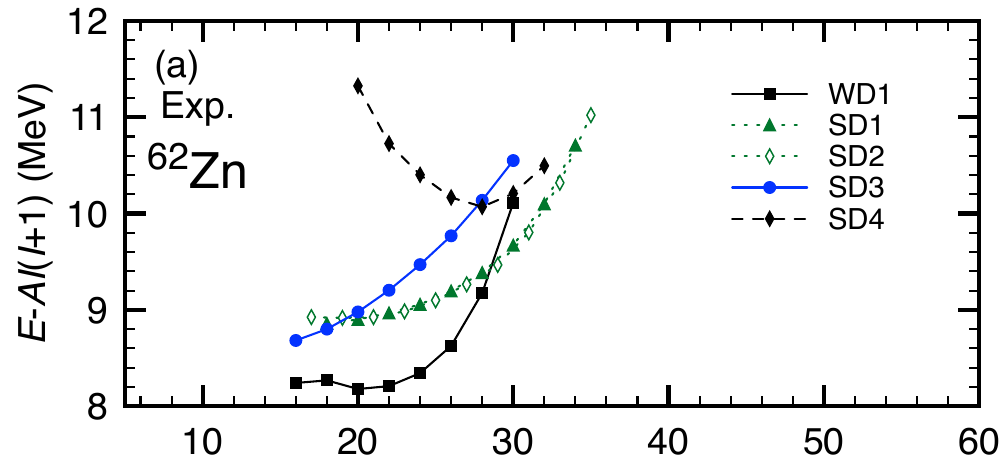}
\includegraphics[scale=0.49]{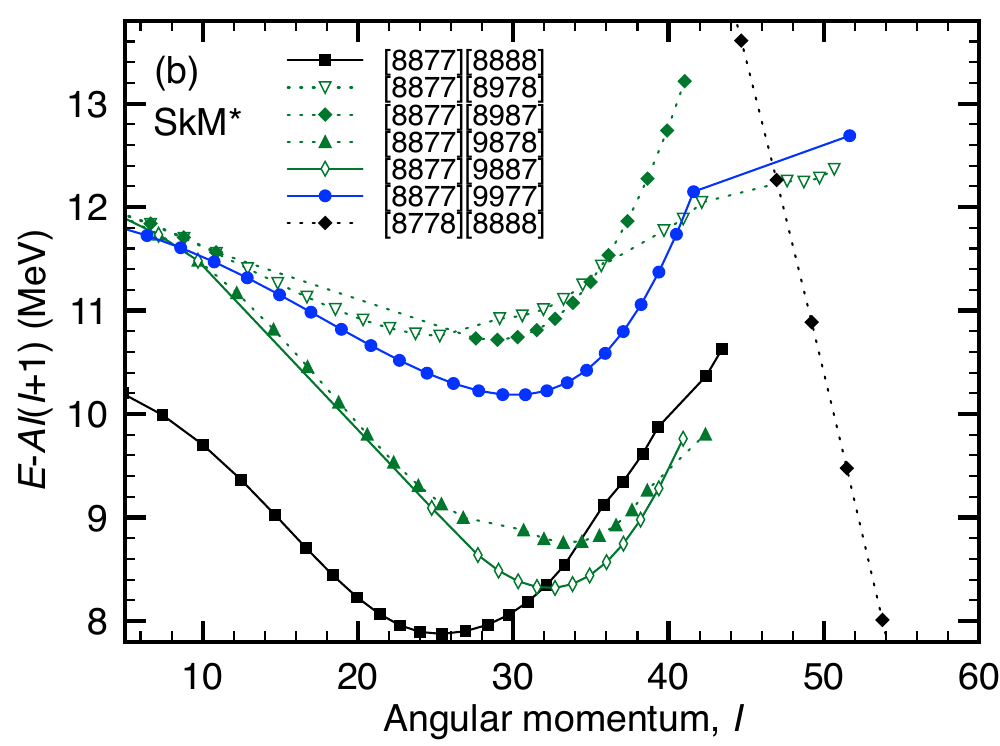}
\caption{\label{fig:62Zn_E}
As Fig.~\ref{fig:60Zn_E} but for $^{62}$Zn. 
(a) Experimental data for the WD1, SD1, SD2, SD3, and SD4 bands~\cite{NNDC,gel12}. 
The parity of SD4 is not defined. 
(b) Numerical results obtained by employing the SkM* functional. 
}
\end{figure}

I show in Fig.~\ref{fig:62Zn_E}(b) the calculated energies for these SD configurations as well as the WD configuration, 
and compare the results with the experimental data~\cite{NNDC} depicted in Fig.~\ref{fig:62Zn_E}(a), 
where SD5 is not drawn because SD2 and SD5 are almost degenerate in energy. 
The authors in Ref.~\cite{mad98} failed to obtain the converged solutions 
for these configurations largely when the SkM* functional was employed.
The cranked-KS equation (\ref{eq:cSKS}) was solved in the 3D mesh in the present calculation, 
whereas the harmonic oscillator basis was adopted in Ref.~\cite{mad98} using {\sc hfodd}~\cite{HFODD}. 

\begin{figure}[t]
\begin{center}
\includegraphics[scale=0.49]{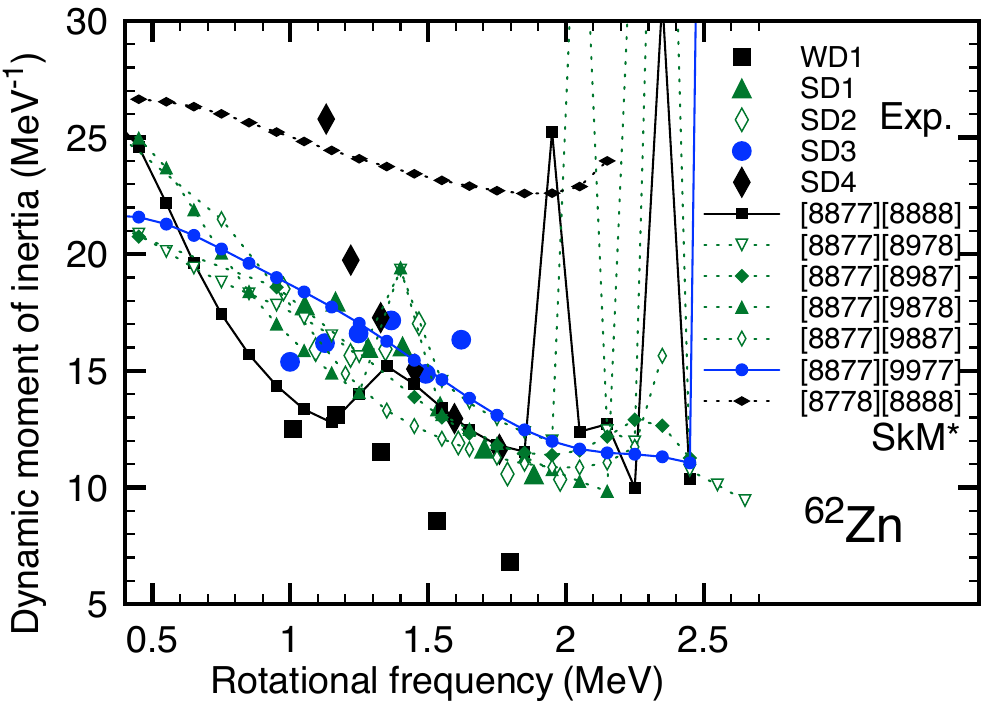}
\caption{\label{fig:62Zn_J2} 
Dynamic moments of inertia plotted as functions of rotational frequency. 
The experimental data~\cite{NNDC} are compared with the numerical results.  
}
\end{center}
\end{figure}

\begin{figure}[t]
\begin{center}
\includegraphics[scale=0.36]{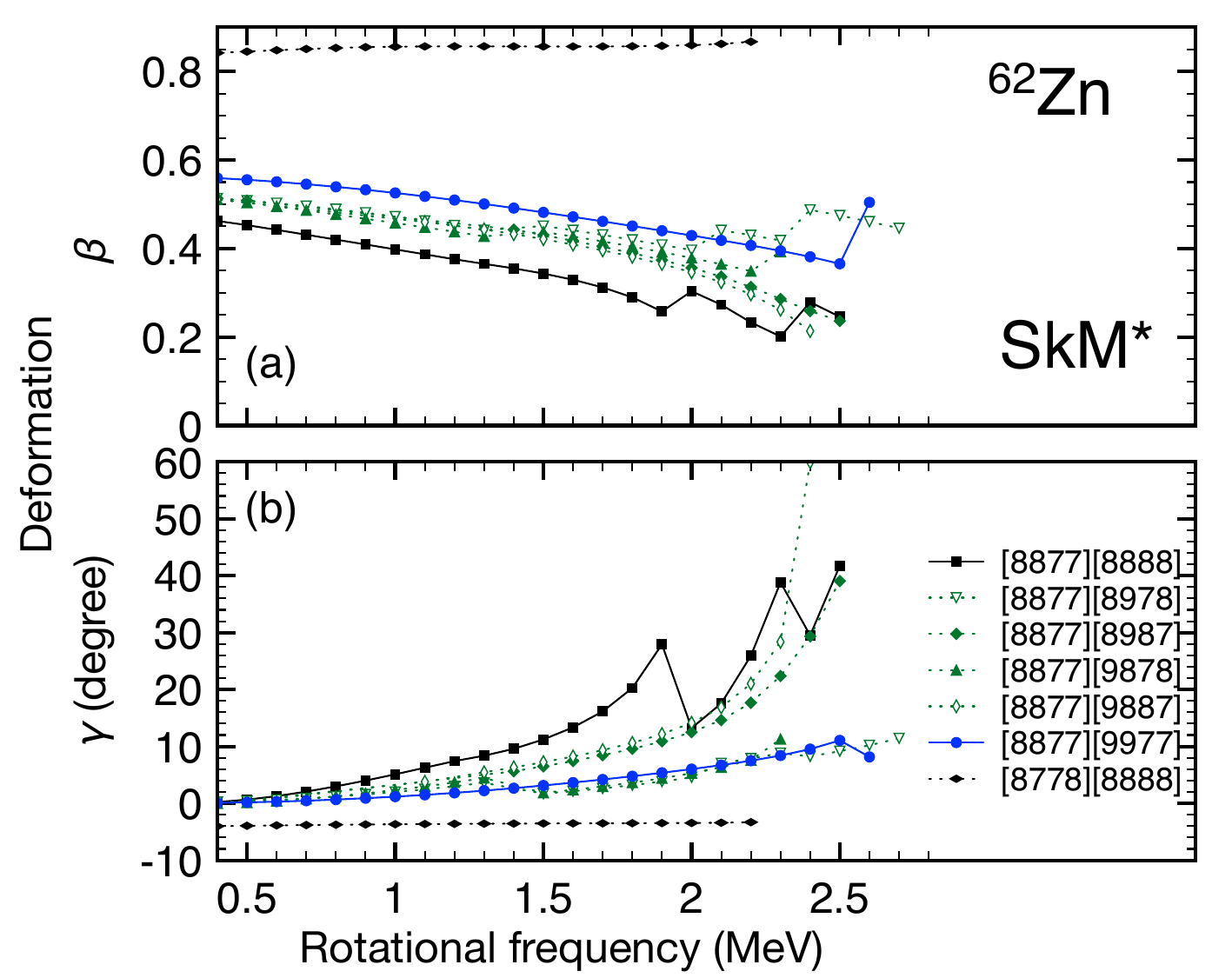}
\caption{\label{fig:62Zn_def} 
Evolution of quadrupole deformation of protons, as a function of the rotational frequency, 
for the selected configurations.  
}
\end{center}
\end{figure}

The WD configuration appears as the yrast band around $I=30$, 
and the negative-parity $[8877]_\pi [9878]_\nu$ and $[8877]_\pi [9887]_\nu$ configurations take over the yrast band above $I \sim 32$. 
This feature seems compatible with the observed energies of WD1, SD1, and SD2 bands. 
However, the calculation does not exactly correspond to the measurement. 
The calculated $[8877]_\pi [8888]_\nu$ configuration continues to increase the spin, 
whereas the maximum spin of the $[22,02]$ configuration is 30.  
In varying the rotational frequency, the structure change occurs; 
the change of the neutron configuration can be seen in the dynamic moments of inertia $\mathcal{J}^{(2)}$ presented in Fig.~\ref{fig:62Zn_J2}.  
One can see a structure change around $\omega_{\rm rot}=1.3$ MeV ($I \sim 26$) in $\mathcal{J}^{(2)}$, 
where the occupied $\alpha=+1/2$ orbital, $[303]7/2$ at $\omega_{\rm rot}=0$ MeV stemming from the $\nu f_{7/2}$ shell, 
crosses with the one stemming from the $pf$ shell.
The occupied $\alpha=-1/2$ orbital stemming from the $\nu f_{7/2}$ shell further 
crosses with a $pf$ orbital at $I=33.3\textrm{--}35.8$ ($\omega_{\rm rot}=1.9\textrm{--}2.0$ MeV). 
Furthermore, this orbital and the one stemming from the $\nu 1h_{11/2}$ shell, $[550]1/2$, 
cross in $I=39.3\textrm{--}42.4$ ($\omega_{\rm rot}=2.3\textrm{--}2.4$ MeV). 
These structural changes are also seen in deformation, as shown in Fig.~\ref{fig:62Zn_def}. 
After the crossing, the deformation $\beta$ increases, and the evolution of triaxiality behaves irregularly. 

In a word, the calculated $[8877]_\pi [8888]_\nu$ configuration, which has a $[22,02]$ character in low spins, 
undergoes crossings with $[22,12]$ around $I=26$, and with $[22,22]$ around $I=34$. 
An additional crossing with $[22,22(1)]$ occurs around $I=40$, where `(1)' stands for one neutron in the $h_{11/2}$ shell. 
Thus, the present calculation overestimates $\mathcal{J}^{(2)}$ of WD1, which is considered as $[22,02]$. 
Note that neither of bands with the configurations $[22,12]$ and $[22,22]$ have been observed experimentally, 
though WD5 is one of the signature-pair bands of $[22,12]$ and terminates at $I=33$~\cite{gel12}. 
\footnote{WD5 is assigned $[22,1(-)2]$ by the calculation with modified Nilsson parameters~\cite{gel14}.}

Then, I discuss the SD configurations. Since the negative-parity bands appear as an SD yrast, 
I first consider the configurations: 
$[8877]_\pi [8978]_\nu$, 
$[8877]_\pi [8987]_\nu$, 
$[8877]_\pi [9878]_\nu$, and  
$[8877]_\pi [9887]_\nu$. 
The signature of neutrons for positive parity is $\alpha=-1/2$ and $\alpha=+1/2$ 
for the former and latter two configurations, respectively.  
As can be seen in Fig.~\ref{fig:62Zn_E}(b), the latter two configurations appear lower in energy 
because the favored orbital $[431]3/2 (\alpha=+1/2)$ is occupied. 
Thus, I am going to investigate these configurations below. 
SD1 and SD2 are experimentally observed as degenerate signature-pair bands, while the degeneracy is broken in the present calculation. 
This is similar to the results of the $[22,23]$ configurations in the CNS approach~\cite{gel12}, 
while the degeneracy is evident for the $[22,13]$ configurations~\cite{gel14}. 

The $[8877]_\pi [9887]_\nu$ configuration appears as the yrast band beyond $I \sim 32$. 
The calculated $\mathcal{J}^{(2)}$ for this configuration reproduces well the measured values of SD2 except for around $\omega_{\rm rot}=1.5$ MeV. 
I find that this configuration becomes oblate at $I=41$ ($\omega_{\rm rot}=2.4$ MeV), as shown in Fig.~\ref{fig:62Zn_def}(b). 
The negative-parity neutron orbitals with $\alpha=+1/2$ near the Fermi level are degenerate, 
thus I could not obtain the converged solutions in the region of $\omega_{\rm rot}=0.5\textrm{--}1.0$ MeV. 
Namely, the $[22,1(-)3(+)]$ and $[22,2(-)3(+)]$ configurations are strongly mixed in the spins $I\lesssim 27$. 
Above this, a $pf$ orbital with $\alpha=+1/2$ is occupied, as in the $[8877]_\pi [8888]_\nu$ configuration. 
Thus, the calculated configuration at $I \gtrsim 27$ may correspond to $[22,2(-)3(+)]$. 

The $[8877]_\pi [9878]_\nu$ configuration appears as a signature-partner band of $[8877]_\pi [9887]_\nu$ beyond $I \sim 32$.
The calculated $\mathcal{J}^{(2)}$ reproduces well the measured values of SD1 for $\omega_{\rm rot}>1.5$ MeV, 
whereas the calculation underestimates the measurement below $\omega_{\rm rot}=1.3$ MeV. 
One finds that the level crossing occurs around $\omega_{\rm rot}=1.4$ MeV. 
A slight increase in deformation can be also seen from $\omega_{\rm rot}=1.3$ MeV to 1.5 MeV in Fig.~\ref{fig:62Zn_def}(a) 
due to the increase in the number of holes in the $f_{7/2}$ shell. 
The eighth and ninth orbitals of a single-particle neutron with ($\mathfrak{p},\alpha$)=($-1,-1/2$) are located close to the Fermi level in low spins. 
Above $\omega_{\rm rot}=1.4$ MeV, a $pf$ shell with $\alpha=-1/2$ is occupied. 
This is why the degeneracy of the calculated signature-pair bands is broken: 
if there is one hole in the $f_{7/2}$ orbital, one can expect the signature-pair bands are degenerate since 
the last $f_{7/2}$ orbitals, [303]7/2, with $\alpha=\pm1/2$ are degenerate even at high spins.  

\begin{figure}[t]
\begin{center}
\includegraphics[scale=0.47]{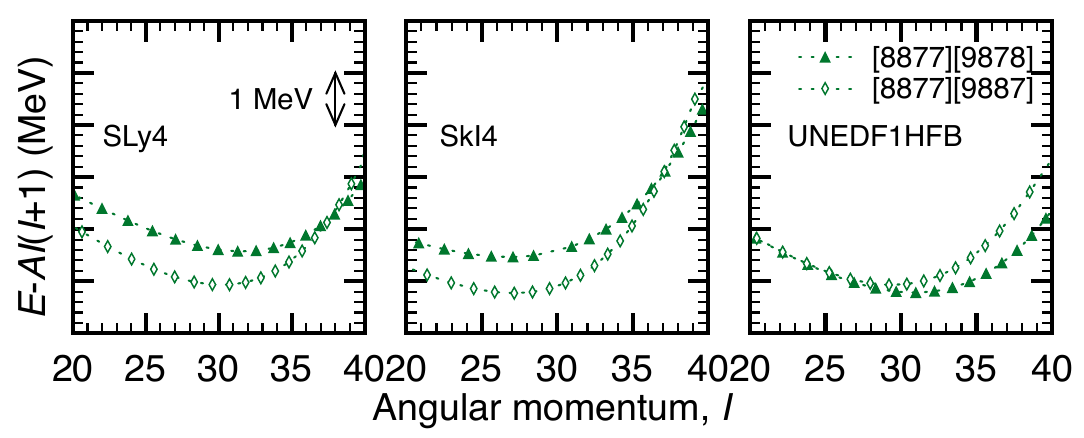}
\caption{\label{fig:62Zn_SD12} 
Calculated relative energies of the $[8877]_\pi [9878]_\nu$ and $[8877]_\pi [9887]_\nu$ configurations employing 
several Skyrme-type functionals.  
}
\end{center}
\end{figure}

One may suspect that breaking of the degeneracy for the $[8877]_\pi [9878]_\nu$ and $[8877]_\pi [9887]_\nu$ configurations 
may depend on the functional one uses. 
Thus, I employ other functionals: SLy4~\cite{cha98}, SkI4~\cite{rei95}, and UNEDF1-HFB~\cite{sch15}. 
In all the calculations, I found two holes in the $\nu f_{7/2}$ shell at $I \gtrsim 20$. 
Then, the degeneracy is broken, as shown in Fig.~\ref{fig:62Zn_SD12}. 
Note that the degeneracy of the signature-pair bands is also broken 
in the calculations employing the relativistic Lagrangians (NL1, NL3, NL-SH, and TM1)~\cite{mad98}.

Another feature I find for the $[8877]_\pi [9878]_\nu$ configuration is that 
the level crossing occurs in $I=38.6\textrm{--}42.4$ ($\omega_{\rm rot}=2.2\textrm{--}2.3$ MeV). 
Accordingly, the deformation parameter $\beta$ increases from 0.35 to 0.39, as shown in Fig.~\ref{fig:62Zn_def}(a). 
Here, the neutron $pf$ orbital with $\alpha=-1/2$ crosses with the one stemming from the $h_{11/2}$ shell. 
Thus the calculated configuration may correspond to $[22,23(1)]$ in Ref.~\cite{gel12}, 
which appears as the yrast band at $I \gtrsim 36$. 

The configurations $[8877]_\pi [8978]_\nu$ and $[8877]_\pi [8987]_\nu$ appears higher in energy 
because of the occupation of the unfavored $\alpha=-1/2$ orbital of the $g_{9/2}$ shell. 
An interesting feature of the $[8877]_\pi [8978]_\nu$ configuration is that 
the deformation increases at some rotational frequencies: 
one sees a sudden increase at $\omega_{\rm rot}=2.1$ MeV and 2.4 MeV. 
This is because the occupied neutron (proton) $pf$ orbital with $\alpha = -1/2$ successively crosses 
with the $h_{11/2}$ shell in $\omega_{\rm rot}=2.0\textrm{--}2.1$ MeV ($I=35.7\textrm{--}39.7$) 
and in $\omega_{\rm rot}=2.3\textrm{--}2.4$ MeV ($I=42.1\textrm{--}47.6$). 

Next, I investigate the positive-parity configuration: two protons and four neutrons occupy the $g_{9/2}$ shell. 
As predicted earlier in Ref.~\cite{mad98}, the positive-parity SD band appears higher in energy. 
The calculated $\mathcal{J}^{(2)}$ for the $[8877]_\pi [9977]_\nu$ configuration reproduces 
the measurement of SD3 reasonably. 
As seen in Fig.~\ref{fig:62Zn_def}(a), the deformation parameter $\beta$ is larger than 
that of the configurations with negative parity. 
One can see that the deformation develops as the occupation number of the $\nu g_{9/2}$ shell increases; 
the deformation parameter $\beta$ around $I \sim 20$ is 0.40, 0.47, and 0.53 
for the WD, negative-parity-SD, and positive-parity-SD state, 
in which two, three, and four neutrons occupy the $g_{9/2}$ shell, respectively. 
The deformation decreases with an increase in spin. 
One finds the crossing in $I=41.6\textrm{--}51.6$ ($\omega_{\rm rot}=2.5\textrm{--}2.6$ MeV). 
This is due to the level crossing with the $\alpha=-1/2$ orbital coming from the $h_{11/2}$ shell. 
Accordingly, the deformation parameter $\beta$ increases from 0.37 to 0.50. 

SD4 has been observed recently, though the parity is not defined yet~\cite{gel12}. 
SD4 exibits high $\mathcal{J}^{(2)}$ values in low $\omega_{\rm rot}$. 
The calculated $\mathcal{J}^{(2)}$ of the $[8877]_\pi [8888]_\nu$ configuration in $\omega_{\rm rot}=1.5\textrm{--}1.8$ MeV 
is comparable to the measurement. 
However, the calculated energies are too low. 
A recent analysis based on the CNS approach suggests SD4 as $[22,22(1)]$~\cite{gel14}. 
In the present calculation, the configuration $[8877]_\pi [8888]_\nu$ corresponds to 
$[22,22(1)]$ above $I \sim 40$. 
Thus, a future experiment at higher spins and the measurement of $\mathcal{J}^{(2)}$ 
will lead to the determination of the configuration of SD4.

\subsection{Superdeformed states in $^{64}$Ge}\label{64Ge}

Various SD states are generated by varying the neutron configuration in $^{62}$Zn. 
In $^{64}$Ge with $N=Z=32$, one can expect the details of the structure information of neutron configurations 
are enhanced thanks to the coherent effect of neutrons and protons. 

\begin{figure*}[t]
\includegraphics[scale=0.49]{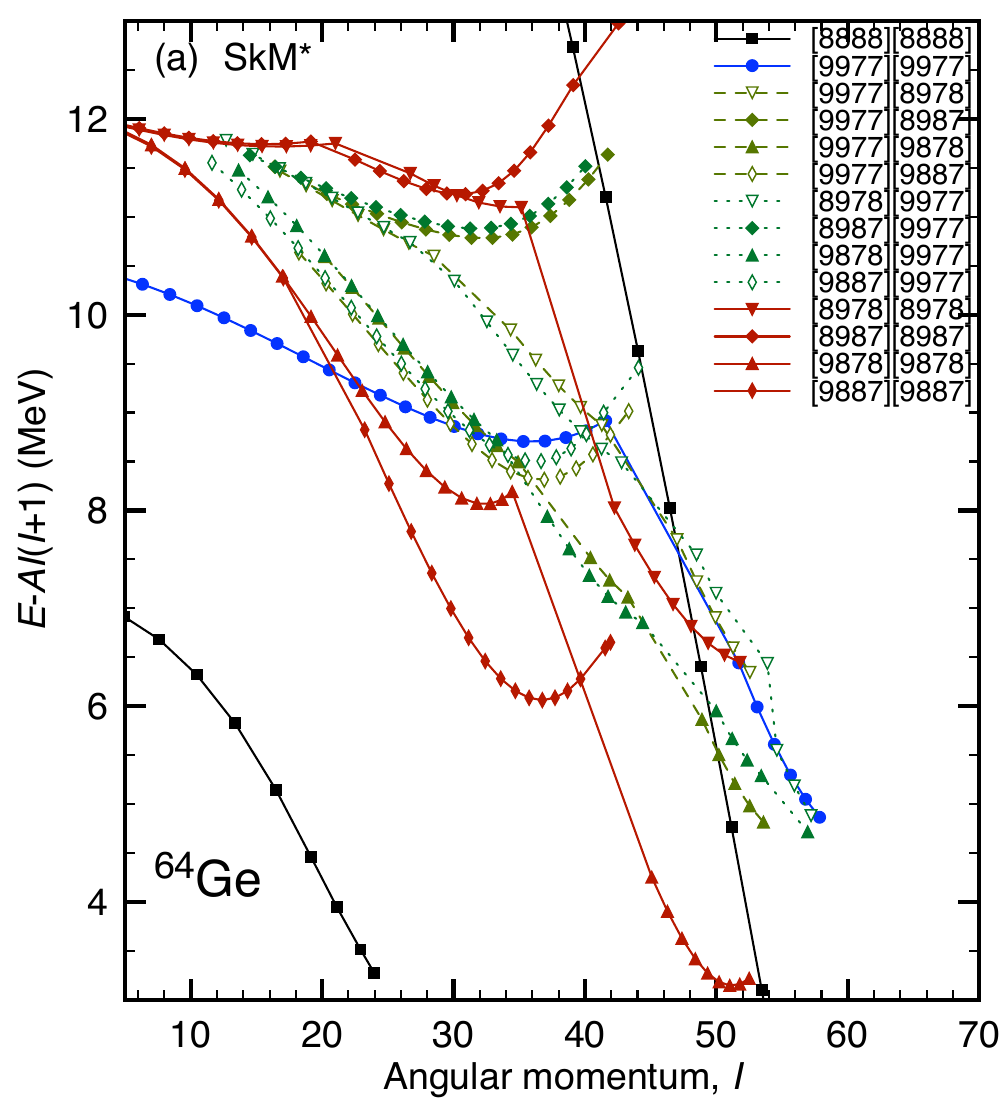}
\includegraphics[scale=0.49]{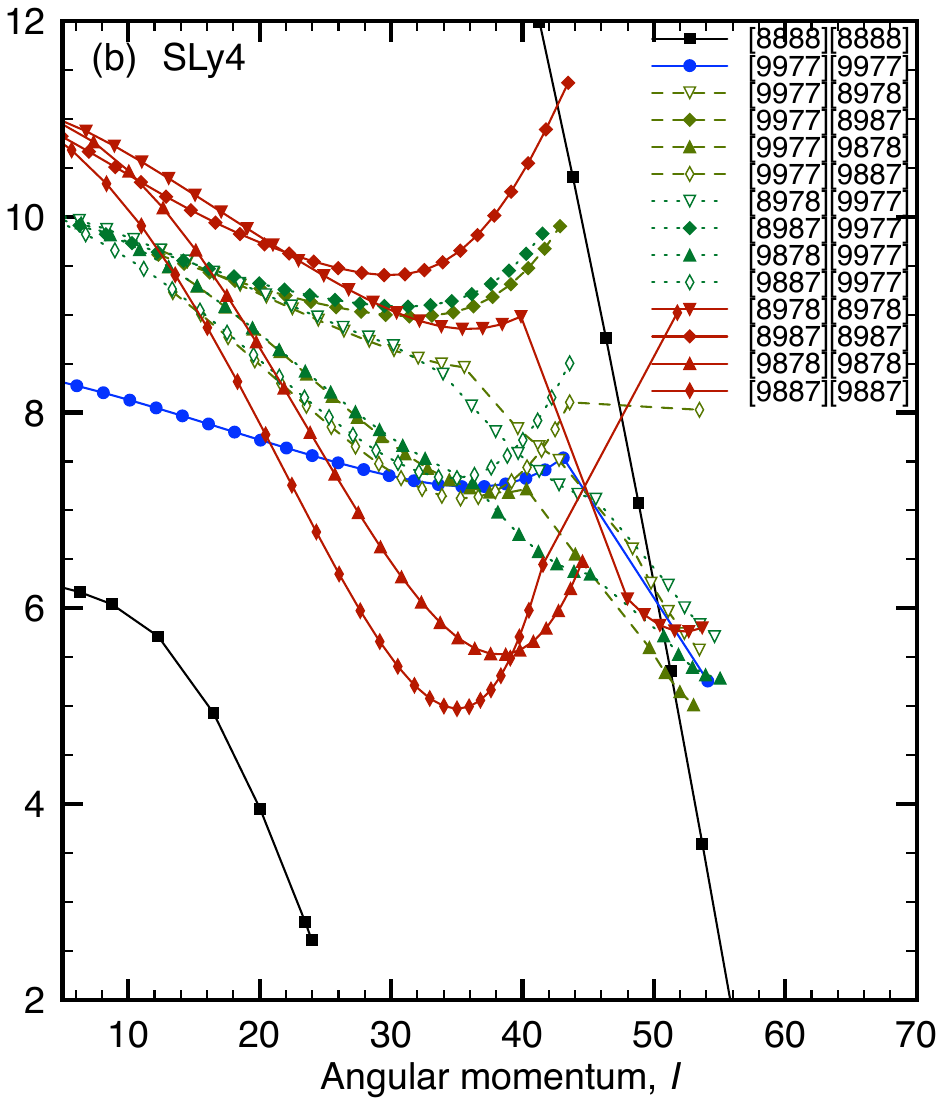}
\caption{\label{fig:64Ge_E}
As Fig.~\ref{fig:62Zn_E} but for $^{64}$Ge, calculated by employing the SkM* and SLy4 functionals.
}
\end{figure*}

\begin{figure}[t]
\begin{center}
\includegraphics[scale=0.65]{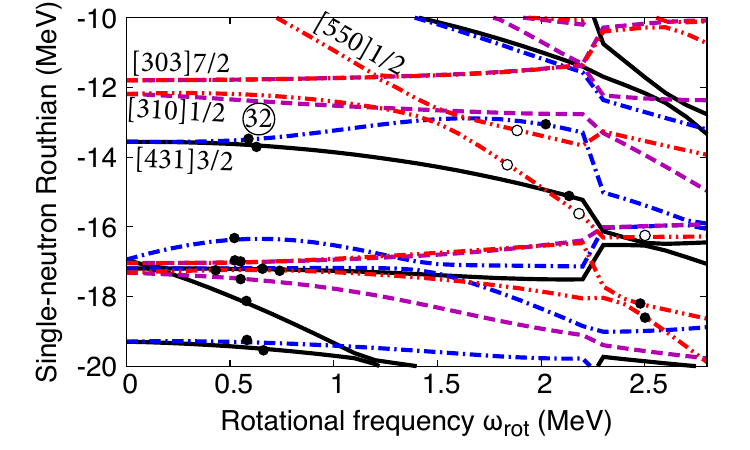}
\caption{\label{fig:64Ge_spe} 
Single-neutron Routhians as functions of the rotational frequency 
for the $[9977]_\pi [9977]_\nu$ configuration in $^{64}$Ge. 
Solid, dash-dotted, dashed, and dash-dot-dotted lines denote the orbitals with $(\mathfrak{p},\alpha)=(+1,+1/2), (+1,-1/2), (-1,+1/2)$, 
and $(-1,-1/2)$, respectively.
}
\end{center}
\end{figure}

\begin{figure}[t]
\begin{center}
\includegraphics[scale=0.36]{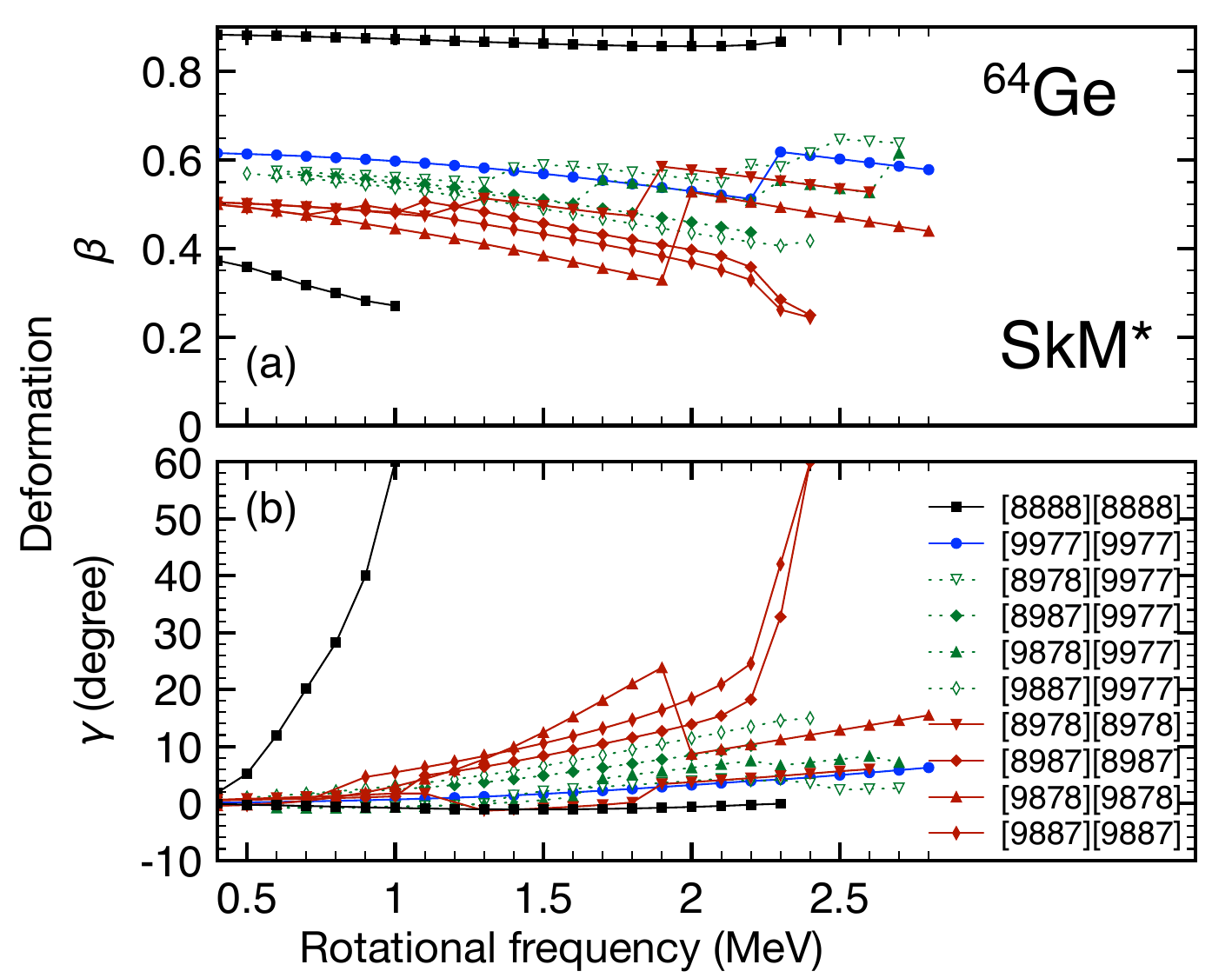}
\caption{\label{fig:64Ge_def} 
As Fig.~\ref{fig:62Zn_def} but for $^{64}$Ge.
}
\end{center}
\end{figure}

I show in Fig.~\ref{fig:64Ge_E} the calculated excitation energies for various configurations. 
To complement the discussion based on the calculation using SkM*, 
the numerical results obtained by employing SLy4 are also displayed. 
The SD configuration $[9977]_\pi [9977]_\nu$, with the quantum number $(\mathfrak{p},\alpha)=(+1,0)$ 
containing four protons and four neutrons 
in the $g_{9/2}$ shell, appears as an SD-yrast band, 
which is in contrast to the SD states in $^{62}$Zn. 
Figure~\ref{fig:64Ge_spe} shows the single-neutron Routhians for this configuration. 
At $\omega_{\rm rot}=0$ MeV, this configuration gives a large deformation: $\beta = 0.62$. 
With an increase in the rotational frequency, deformation is weakened, as seen in $^{62}$Zn. 
One finds in Fig.~\ref{fig:64Ge_spe} that the occupied $pf$ orbital with $\alpha=-1/2$ 
crosses with the one coming down from the $h_{11/2}$ orbital in $\omega_{\rm rot}=2.2\text{--}2.3$ MeV. 
I find also the level crossing in protons. 
Consequently, the $[9977]_\pi [9977]_\nu$ configuration beyond $I \sim 52$ possesses one proton and one neutron 
in the $h_{11/2}$ shell, and the deformation develops from $\beta=0.51$ to 0.62, as shown in Fig.~\ref{fig:64Ge_def}(a). 
A similar crossing occurs in $\omega_{\rm rot}=2.3\text{--}2.5$ MeV ($I=43.1\textrm{--}54.1$) in the calculation with SLy4.

As shown in Fig.~\ref{fig:64Ge_spe}, the single-particle orbitals just below and above the SD gap at $N,Z=32$ are 
$[431]3/2$ and $[310]1/2$, respectively, for both protons and neutrons at $\omega_{\rm rot}=0$ MeV.
The ``1p1h'' and ``2p2h'' excitations 
from the positive-parity SD configuration may appear near the yrast line. 
I then investigate these configurations. 

There are four proton ``1p1h'' excitations from $[431]3/2(\alpha=\pm 1/2)$ to 
$[310]1/2(\alpha=\pm 1/2)$, labeled as $[8978]_\pi, [8987]_\pi, [9878]_\pi$, and $[9887]_\pi$,
while keeping the neutron configuration $[9977]_\nu$. 
They are negative parity and correspond to the configuration involving three protons and four neutrons in the $g_{9/2}$ shell. 
Calculated results for these configurations
are shown in Figs.~\ref{fig:64Ge_E}, \ref{fig:64Ge_def}. 
For all the four negative-parity configurations 
I obtain solutions that commonly have large deformation $\beta\sim 0.5$ at low
$\omega_{\rm rot} \lesssim 1.3$ MeV ($I \sim 26$). 
At higher spins, however, a difference grows up. 
Nevertheless, I observe the following systematic trends.

The deformation of 
the configuration $[8987]_\pi [9977]_\nu$ with quantum number $(\mathfrak{p},\alpha)=(-1,0)$, 
i.e. a particle--hole excitation of $\pi[431]3/2 (\alpha=+1/2) \to \pi[310]1/2 (\alpha=+1/2)$ across the SD gap, 
decreases monotonically as increasing spin or rotational frequency. 
Then, the upbend appears in Fig.~\ref{fig:64Ge_E} at high spins. 
The deformation of 
the configuration $[9887]_\pi [9977]_\nu$ with $(\mathfrak{p},\alpha)=(-1,1)$, 
another particle--hole excitation of $\pi[431]3/2 (\alpha=-1/2) \to \pi[310]1/2 (\alpha=+1/2)$, 
also decreases with an increase in spin. 
For both configurations, a single-particle proton is promoted to the orbital with $\alpha=+1/2$. 
On the other hand, 
the deformation of $[8978]_\pi [9977]_\nu$ with $(\mathfrak{p},\alpha)=(-1,1)$ 
increases successively at $\omega_{\rm rot}=1.4$ MeV and 2.2 MeV. 
This is because the $\pi[310]1/2 (\alpha=-1/2)$ orbital crosses with the $\pi[550]1/2 (\alpha=-1/2)$ orbital 
in $\omega_{\rm rot}=1.3\textrm{--}1.4$ MeV ($I=26.7\textrm{--}30$), 
and then the neutron occupied $pf$ orbital with $\alpha=-1/2$ and the $\nu[550]1/2 (\alpha=-1/2)$ orbital 
cross in $\omega_{\rm rot}=2.1\textrm{--}2.2$ MeV ($I=42.8\textrm{--}48.5$). 
The deformation of $[9878]_\pi [9977]_\nu$ with $(\mathfrak{p},\alpha)=(-1,0)$ also 
increases successively at $\omega_{\rm rot}=1.7$ MeV ($I = 37.1$) and 
2.3 MeV ($I=50$) due to the crossings similarly. 
In these two configurations, a single-particle proton is promoted to the orbital with $\alpha=-1/2$, 
same as the $[8877]_\pi [8978]_\nu$ and $[8877]_\pi [9878]_\nu$ configurations in $^{62}$Zn.  

The calculated negative-parity signature-pair bands appearing in low energies in $^{62}$Zn, 
$[8877]_\pi [9878]_\nu$ and $[8877]_\pi [9887]_\nu$, exhibited a weak signature splitting. 
Similarly in $^{64}$Ge here, the configurations $[9878]_\pi [9977]_\nu$ and $[9887]_\pi [9977]_\nu$ appear as 
signature-pair bands up to the level crossings. 
The configurations $[9878]$ and $[9887]$ correspond to the particle--hole excitation 
from the unfavored $[431]1/2 (\alpha=-1/2)$ orbital to the $[310]1/2 (\alpha=\pm 1/2)$ orbital. 
As seen in Fig.~\ref{fig:64Ge_spe}, the particle--hole excitation energy decreases as increasing the rotational frequency. 
The signature splitting occurs because the single-particle proton is promoted to the $[310]1/2$ orbital. 
If the energy ordering of the $[303]7/2$ and $[310]1/2$ orbitals is interchanged, 
one can expect the degeneracy of the negative-parity signature-pair bands in $^{64}$Ge as in the case of $^{62}$Zn. 

I mention here that in the present study, I obtained the neutron ``1p1h'' excited states, i.e. 
$[9977]_\pi [8978]_\nu$ and so on, which appear close to the 
proton particle--hole-excited states as in $^{60}$Zn because the system under consideration is an $N=Z$ nucleus.
The negative-parity ``1p1h'' states appear at about 1.8 MeV 
above the positive-parity SD state at $\omega_{\rm rot}=0$ MeV, 
and the positive-parity SD band is yrast up to $I \sim 22$.
This SD-shell closure is not as strong as in $^{60}$Zn, where $\sim 4$ MeV gap energy is seen.
However, this is compatible with the gap in an SD doubly-magic $^{40}$Ca nucleus: 1.8 MeV with SkM*~\cite{sak20} and $\sim 2.2$ MeV with NL3*~\cite{ray16}.

Finally, I consider ``2p2h'' excitations with respect to the positive-parity SD configuration, 
which have ``1p1h'' excitations $[431]3/2 \to [310]1/2$ both in protons and neutrons, 
and hence positive parity. 
Some of them actually appear as yrast bands at the intermediate spins, as shown in Fig.~\ref{fig:64Ge_E}. 
The deformation of these configurations are, however, weaker than those of the ``1p1h'' excited states, 
and the deformation parameter is around $\beta \sim 0.5$, 
which is in between the deformation of the WD $[8888]_\pi [8888]_\nu$ and SD $[9977]_\pi [9977]_\nu$ configurations, 
as shown in Fig.~\ref{fig:64Ge_def}(a). 
This is because a ``2p2h'' excitation on the positive-parity SD configuration can also be regarded as a 
``2p2h'' excitation on the WD configuration. 
Namely, the configurations involve three protons and three neutrons in the $g_{9/2}$ shell. 
Since the single-particle proton and neutron both occupy the favored $\alpha=+1/2$ orbital of the $g_{9/2}$ shell, 
the configurations $[9878]_\pi [9878]_\nu$ and $[9887]_\pi [9887]_\nu$ appear low in energy. 
An interesting feature is that the configurations $[8978]_\pi [8978]_\nu$ and $[9878]_\pi [9878]_\nu$ undergo  
the crossing around $I=35$ ($I=40$) with SkM* (SLy4), 
though I could not trace the configuration $[9878]_\pi [9878]_\nu$ in higher spins with SLy4. 
In these configurations, a single-particle proton and neutron are promoted to the orbital with $\alpha=-1/2$. 
After the crossing, the configuration involves one proton and one neutron in the $h_{11/2}$ shell. 
Accordingly, deformation develops.

At the end of this section, I reiterate and summarize what I have discussed concerning the SD states.
A distinct feature of the positive-parity SD band in $^{64}$Ge from that in $^{62}$Zn 
is that the positive-parity SD band in $^{64}$Ge appears lower in energy than the  ``1p1h'' and ``2p2h'' excited 
bands in $I \lesssim 20$. 
It should also be mentioned that the deformation parameters for the positive-parity SD band in $^{64}$Ge are 
larger than those in $^{62}$Zn. 
Although the particle number $30$ is a strong SD magic number, this appears in a smaller deformation region. 
Actually, the SkM* functional produces $\beta=0.47$ at $I\sim20$ ($\omega_{\rm rot}=0.9$ MeV) for the positive-parity SD band in $^{60}$Zn, 
whereas the calculated $\beta$ is 0.60 at $I \sim 20$ ($\omega_{\rm rot}=1.0$ MeV) in $^{64}$Ge. 
The competing shell effect of protons and neutrons, 
the former prefers a weaker deformation than the latter, 
makes the SD structures convoluted in $^{62}$Zn. 
The SD-shell gap at 32 appears in a larger deformation region than 30 in the deformed Woods-Saxon and 
the Nilsson potentials as a manifestation of the approximate pseudo-SU(3) symmetry~\cite{dud87}. 
This finding indicates that 
the selfconsistent mean-field, or the Kohn--Sham potential, also obeys the approximate pseudo-SU(3) symmetry. 
However, the SD-shell closure is much stronger at 30 than at 32.

\subsection{Hyperdeformation in $^{60,62}$Zn and $^{64}$Ge}\label{hyper}

\begin{figure}[t]
\begin{center}
\includegraphics[scale=0.65]{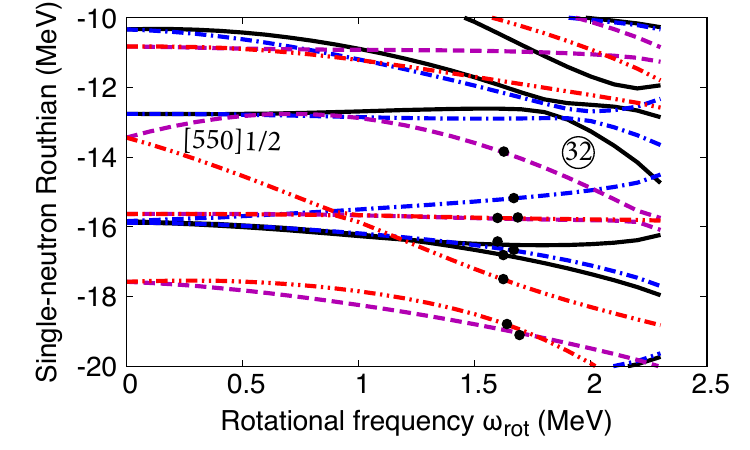}
\caption{\label{fig:64Ge_HD_spe} 
As Fig.~\ref{fig:64Ge_spe} but for the HD $[8888]_\pi [8888]_\nu$ configuration. 
}
\end{center}
\end{figure}

\begin{figure}[t]
\begin{center}
\includegraphics[scale=0.42]{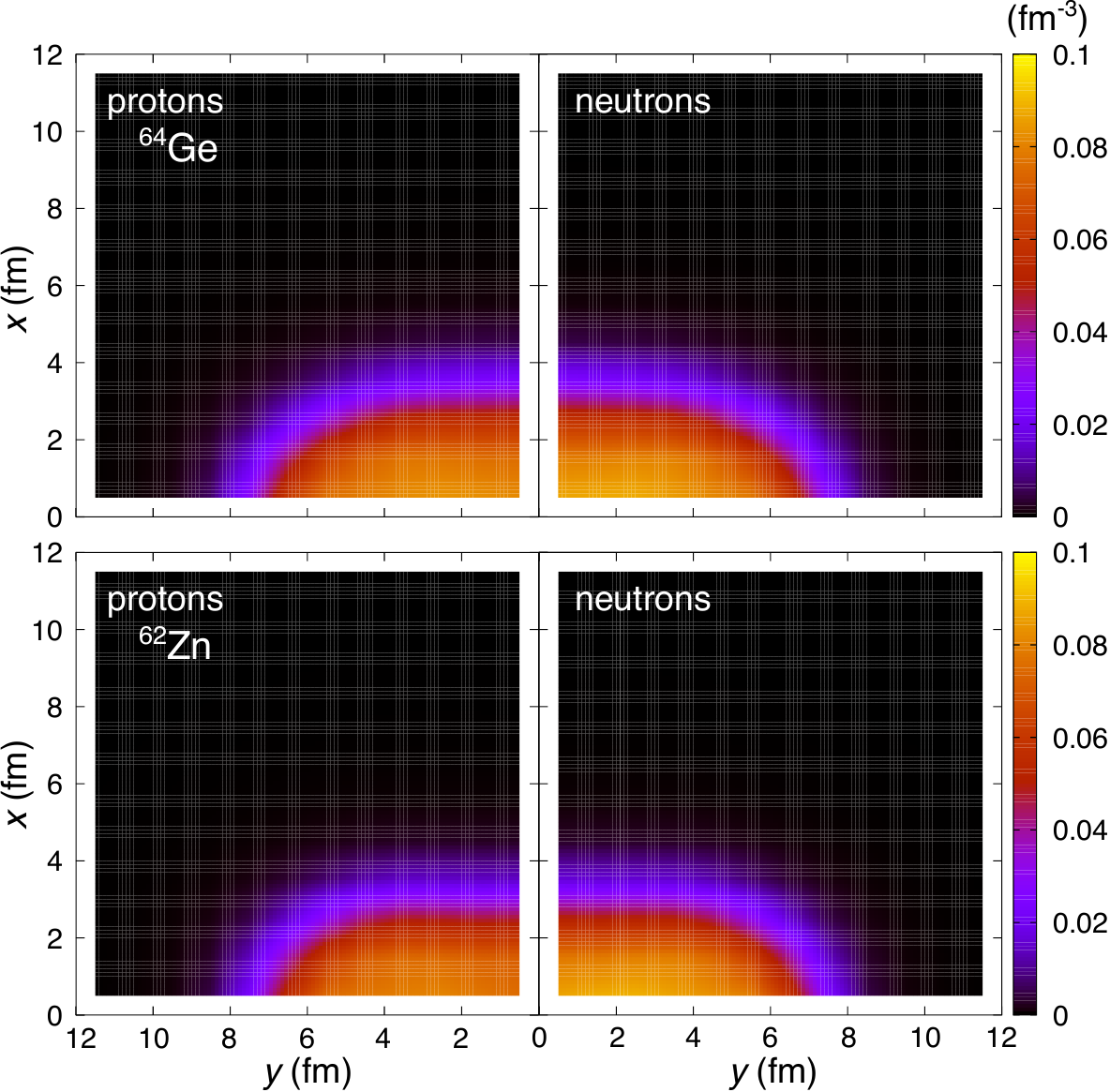}
\caption{\label{fig:HD} 
Particle density distributions for protons (left) and neutrons (right) 
of the HD state in $^{64}$Ge (upper) and $^{62}$Zn (lower) at $\omega_{\rm rot}=2.0$ MeV. 
}
\end{center}
\end{figure}

I investigate the possible occurrence of the hyperdeformed (HD) states in $^{60,62}$Zn and $^{64}$Ge. 
First, I discuss the HD states in $^{64}$Ge, in which two protons and two neutrons 
are promoted in the $h_{11/2}$ shell. 
Thus, I show in Fig.~\ref{fig:64Ge_HD_spe} the single-particle Routhians of neutrons for the $[8888]_\pi [8888]_\nu$ configuration 
with the use of the SkM* functional. 
The single-proton Routhians are basically the same as neutrons. 
One sees that the $[550]1/2$ orbitals stemming from the $h_{11/2}$ shell are occupied. 
Remarkably, an HD-shell gap at $N, Z=32$ emerges around $\omega_{\rm rot}=2.0$ MeV. 
It should be noticed that the early theoretical calculation based on the Woods-Saxon potential 
predicts the enhanced HD-shell gap of 32 at $I=0$~\cite{dud88}. 
In the present calculation employing the SkM* functional, however, 
the particle number 30 gives a higher energy gap than 32 at $\omega_{\rm rot}=0$ MeV. 
One can expect the HD band is yrast at high spins. 
Indeed, I find the yrast band at $I \gtrsim 50$, as seen in Fig.~\ref{fig:64Ge_E}. 
The deformation parameter is constantly large: $\beta=0.85\textrm{--}0.89$. 
The density distributions for protons and neutrons at $\omega_{\rm rot}=2.0$ MeV ($I=53.5$) 
are displayed in the upper part of Fig.~\ref{fig:HD}. 
The rms matter size for the shortest axis $\sqrt{\langle x^2\rangle}$ is 1.75 fm, while for the longest axis $\sqrt{\langle y^2\rangle}=3.76$ fm. 
The ratio exceeds 2.1.

One sees in Fig.~\ref{fig:64Ge_HD_spe} that there appears an HD gap at 30 around $\omega_{\rm rot}=2.0$ MeV. 
Thus, I investigate 
the corresponding $[8778]_\pi [8778]_\nu$ configuration in $^{60}$Zn and 
the $[8778]_\pi [8888]_\nu$ configuration in $^{62}$Zn. 
The energies are shown in Figs.~\ref{fig:60Zn_E}, \ref{fig:62Zn_E}. 
The dynamic moments of inertia are much greater than those of the SD states, 
and the deformation parameter is almost constant $\beta=0.83\textrm{--}0.86$. 
Although the configuration involves two neutrons and only one proton in the $h_{11/2}$ shell, 
the density distributions of the HD state in $^{62}$Zn at $\omega_{\rm rot}=2.0$ MeV 
shown in the lower panel of Fig.~\ref{fig:HD} are compatible with those of the HD state in $^{64}$Ge. 

\section{Summary}\label{summary}

I have investigated the highly-deformed states at high spins in even-even $N \simeq Z$ nuclei with $A\sim 60$.
The Skyrme EDFs were applied to describe various near-yrast structures  
in the framework of the configuration-constrained cranked-KS. 
The KS orbitals are represented in the Cartesian mesh to depict both the SD and HD shapes flexibly.

The positive-parity SD band involving two protons and two neutrons in the intruder $g_{9/2}$ shell 
appears far below the negative-parity SD bands in $^{60}$Zn, confirming the doubly-magic SD structure. 
Adding two neutrons causes the SD states to be complicated because the neutron number 32 
favors stronger deformation than 30. 
The negative-parity SD bands involving two protons and three neutrons in the $g_{9/2}$ shell 
appear as the SD-yrast bands, whereas the positive-parity SD band involving one more neutron in the $g_{9/2}$ shell 
emerges higher in energy. 
I have found that the positive-parity SD band in $^{64}$Ge involving two more protons in the $g_{9/2}$ shell 
appears as an SD-yrast band, and 
that it is strongly deformed compared with the positive-parity SD bands in $^{60,62}$Zn. 

In the course of the systematic investigation, 
I have found the HD states in rapidly rotating $^{60}$Zn and $^{64}$Ge at $\omega_{\rm rot}\sim 2.0$ MeV. 
The present calculation predicts that the positive-parity HD states appear 
as the yrast band at $I \gtrsim 50$ in $^{60}$Zn and $^{64}$Ge, 
while the negative-parity even-spin states appear as the yrast band in $^{62}$Zn.

\begin{acknowledgments} 
This work was supported by the JSPS KAKENHI (Grants No. JP19K03824 and No. JP19K03872). 
The numerical calculations were performed on the computing facilities  
at the Yukawa Institute for Theoretical Physics, Kyoto University, 
and at the Research Center for Nuclear Physics, Osaka University. 

\end{acknowledgments}

%

\end{document}